\documentclass[runningheads]{llncs}
\usepackage{graphicx}
\usepackage{tcolorbox}

\usepackage{xspace}
\usepackage{tabularx}
\usepackage{subcaption}
\usepackage{graphicx}

\newcommand{\DQ}{DQ\xspace}
\newcommand{\DQStart}{DQ\xspace}

\begin{document}
\title{On the Automated Processing of User Feedback}
\titlerunning{Automated User Feedback Analysis}
%
\author{Walid Maalej\inst{1}\and 
Volodymyr Biryuk\inst{1}\and
Jialiang Wei\inst{2}\and
Fabian Panse\inst{3} 
}
\authorrunning{Maalej et al.}
%
\institute{University of Hamburg, Department of Informatics, Hamburg 22527, Germany \\
\email{\{walid.maalej,volodymyr.biryuk\}@uni-hamburg.de} \and
EuroMov Digital Health in Motion, Univ Montpellier, IMT Mines Alès,  30100 Alès, France \\
\email{jialiang.wei@mines-ales.fr} \and
Hasso-Plattner-Institut für Digital Engineering gGmbH, 14482 Potsdam, Germany \\
\email{fabian.panse@hpi.de} 
}

\maketitle              
\begin{abstract} 
User feedback is becoming an increasingly important source of information for requirements engineering, user interface design, and software engineering in general. 
Nowadays, user feedback is largely available and easily accessible in social media, product forums, or app stores. 
Over the last decade, research has shown that user feedback can help software teams:  
a) better understand how users are actually using specific product features and components, 
b) faster identify, reproduce, and fix defects, and 
b) get inspirations for improvements or new features. 

However, to tap the full potential of feedback, there are two main challenges that need to be solved.
First, software vendors must cope with a large quantity of feedback data, which is hard to manage manually. 
Second, vendors must also cope with a varying quality of feedback as some items might be uninformative, repetitive, or simply wrong.
This chapter summarises and pipelines various data mining, machine learning, and natural language processing techniques, including recent Large Language Models, to cope with the quantity and quality challenges. 
We guide researchers and practitioners through  implementing effective, actionable analysis of user feedback for software and requirements engineering.

\end{abstract}

\section{Introduction}

Traditionally, software teams collect and analyse requirements in early project phases by involving (some) users through questionnaires, interviews, or workshops. 
This has dramatically changed with the emergence of app stores as social-media-like software distribution platforms and the emergence of agile as incremental development methodology. 
Over the last decade, requirements engineering has become more dynamic, spanning the entire product lifecycle, and regularly taking into account the voices of (many) users. 
User satisfaction and user feedback nowadays represent an important input for developers and vendors when deciding what to develop and what to release next \cite{maalejDataDrivenRequirementsEngineering2016,Li:ICSE:2024}.

One common form of feedback are app reviews.  With the rise of distribution platforms like Apple's App Store and Google Play user-developer communication \cite{Hassan:EMSE:2018} has become not only much easier but also more crucial. 
Users can easily submit their opinions, inquiries, and requests via app reviews. 
Such feedback might include important information for software teams, such as how a new feature is perceived and used or that a User Interface (UI) element does not behave as expected in a certain context. 
This feedback is also publicly visible to other users and  competitors and might thus impact the app success. 
Ignoring it can lead in the worst case to the fall of even popular apps \cite{Williams:RE:18,Martens:Software:19}. 

Comments in social media platforms such as Twitter or Reddit represent another popular form of user feedback. 
The professional analysis of these comments for insights gained popularity across various domains---not only in politics, business, and education, but also in software and technology \cite{Reimer:DJ:2023}. 
Communities centred around specific software tools or topics have emerged, with similar dynamics like app reviews: i.e.,~regularly discussing software and how it does or does not supply the needs of the community. 
Some software vendors provide a substantial  support to their users in social media, by responding to inquiries, reacting to suggestions, and informing about new updates \cite{Martens:RE:2019}.

While these trends created an opportunity for software vendors to monitor how users perceive newly released features and to react accordingly, the trends also created two major challenges \cite{Li:ICSE:2024,Reimer:DJ:2023}. 
First, it becomes hard and impractical to cope with the sheer \textit{feedback quantity}. 
Particularly popular apps might receive thousands of feedback entries per day \cite{Pagano2013} in different feedback channels. 
Second, the fact that ``anyone can submit  anything'' naturally leads to diverging \textit{feedback quality}. 
While some feedback includes detailed and actionable information such as a description of a defect, the context in which it occurs, and the steps to reproduce it; other feedback is ambiguous, hardly readable, too verbose, or simply wrong \cite{Martens:EMSE:2019,Martens:RE:2019}. 

In recent years, researchers and startups have suggested numerous feedback processing approaches, particularly for coping with the feedback quantity challenge. 
Most notably, these approaches often aim to \textit{classify or cluster} feedback according to its subject and the information it includes. 
Generally, a feedback item can either be classified vertically as a bug reports, feature request, user experience, inquiry etc.~or horizontally according to commonly discussed features of the app, certain app issues, or app components. 
Other approaches from the literature also aim to \textit{summarise}  feedback items and the users' sentiments as well as to \textit{match} certain feedback to development artefacts such as tasks in the issue tracker which might be affected. 
With the recent rise of Large Language Models (LLMs) the accuracy of feedback processing (particularly classification, summarisation, and sentiment analysis) is reaching a top level, which will certainly further increase the applicability of feedback analysis in practice. 
Finally, researchers also recently started to investigate the feedback quality challenge: predicting the feedback authenticity, collecting additional contextual information to improve the feedback understandability and actionability, or normalising feedback with regard to the user demographics and other contextual factors. 

This chapter overviews the field of automated feedback processing---a rather convoluted and fragmented field in the meantime. 
We discuss and structure various approaches from literature and highlight pitfalls and open issues: hoping to guide practitioners and researchers through the field and to  implement effective feedback processing for software and requirements engineering.
Section \ref{sec:usecases} presents four main use cases of user feedback analysis covering various design and engineering tasks.
Section \ref{sec:pipeline} presents a feedback processing pipeline, discussing state-of-the-art approaches to preprocess, augment, classify, cluster, and summarise user feedback as well as to match it with software engineering artefacts. 
Then, Section \ref{sec:quality} focuses on feedback quality by discussing how data quality management techniques can potentially be applied for user feedback. 
Finally, Section \ref{sec:discussion} concludes the chapter and highlights future directions.   
The chapter is accompanied by several computational notebooks, partly compiled from previous work. Readers can thus try out the feedback processing techniques and use cases~\cite{maalej2024z}.

\section{Usefulness of User Feedback}
\label{sec:usecases}
The usefulness of feedback processing for requirements engineering and software engineering has been discussed through a variety of use cases targeting various roles: from requirements analysts, product owners, and UX experts to quality engineers, developers, and project managers. 
We discuss the feedback usefulness  along four main use cases: understanding user needs, monitoring and improving quality, planning and scoping, as well as supporting users.

\begin{figure}[]
\centering
\begin{subfigure}{.5\textwidth}
  \centering
  \includegraphics[width=.45\linewidth]{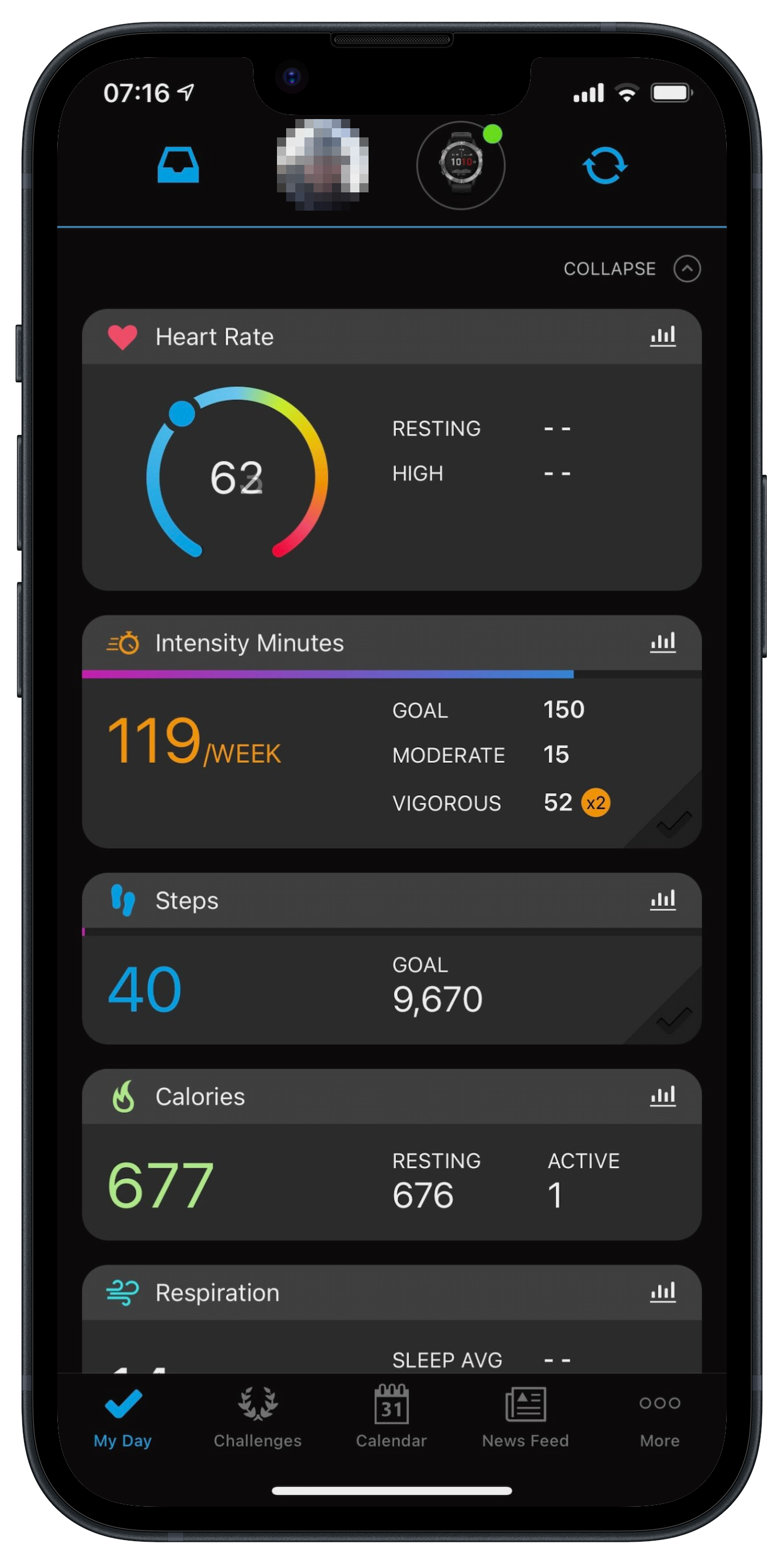}
  \caption{Garmin connect (health monitoring)}
  \label{fig:sub1}
\end{subfigure}%
\begin{subfigure}{.5\textwidth}
  \centering
  \includegraphics[width=.45\linewidth]{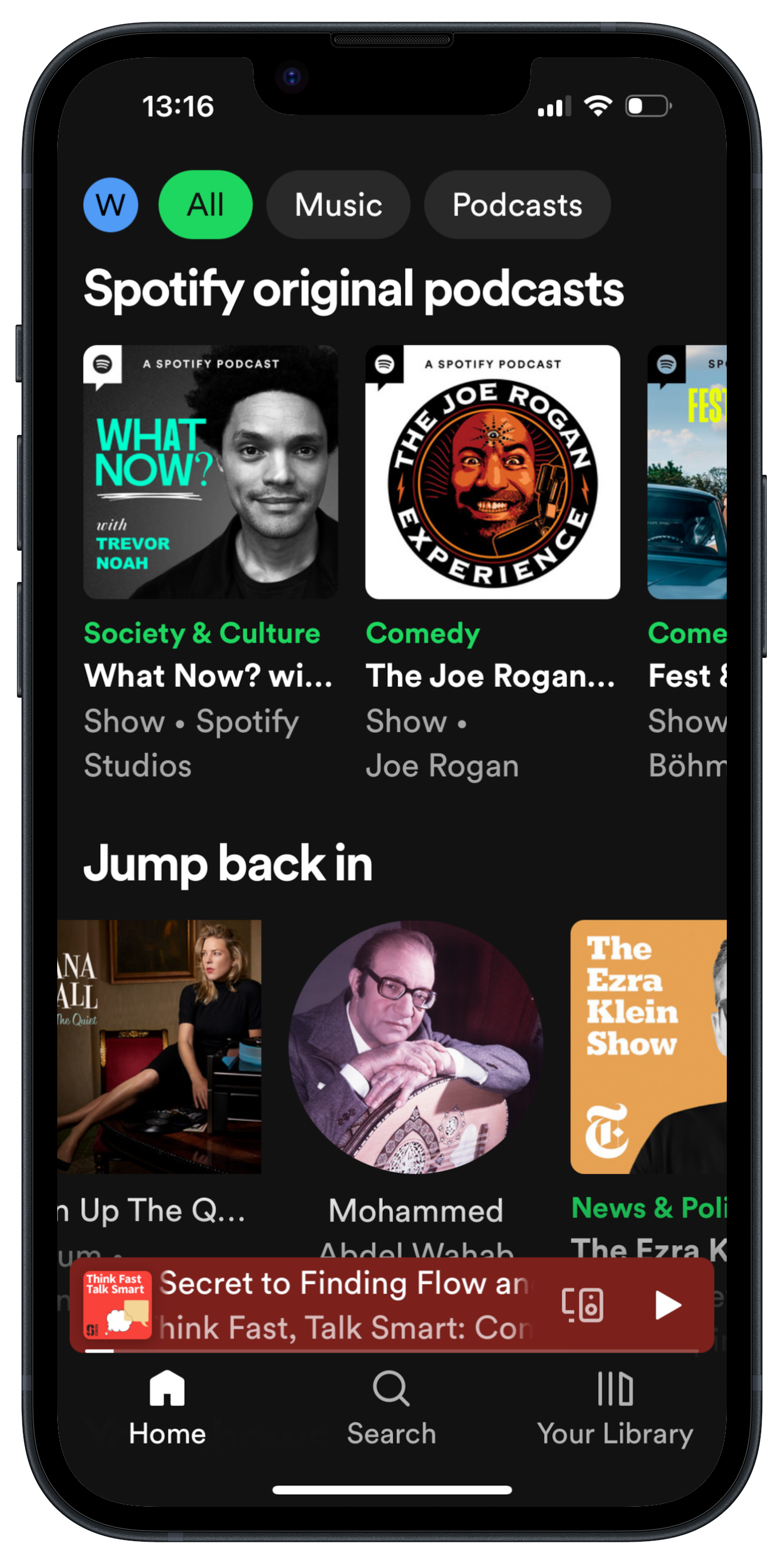}
  \caption{Spotify (audio streaming)}
  \label{fig:sub2}
\end{subfigure}
\caption{Screenshots for two example apps to discuss user feedback processing.}
\label{fig:running-examples}
\end{figure}

We use two example apps as shown on Figure~\ref{fig:running-examples}. 
The first, Garmin Connect, refers to health and fitness monitoring apps. 
Compatible health monitoring devices can be paired with the phone, enabling to track user activities as ruining, swimming, lifting weights, etc.
The app provides detailed reports of the user health data, conducts in-depth analysis of workouts, and creates personalised training plans. 
Users can also assess their workout progress compared to others and achieve badges.
The second example, Spotify, refers to streaming apps. 
Users can play music tracks, podcasts, or audiobooks, which may also include a display of lyrics or video clips. 
Users can register to particular artists, genres, or  channels to get new content. 
They can compile playlists and share them with others. 
Finally, users can rate tracks, playlists, or artists and get recommendations according to their preferences.

\subsection{Understanding User Needs}
Generally, monitoring trends on social media (for a specific app or a category of apps) helps understand emergent topics, user needs, and how they change over time \cite{Martens:Software:19}.
In particular, some user may request new features or new content \cite{Pagano2013}.
Users might, for instance, describe a feature they miss like: 
``I would like to see my Spotify playlists shown in Garmin connect and get played when I start my workouts''. 
Or, users might ask: ``please consider adding a visualisation of my sleeping/standing/sitting ratio. 
A balanced ratio is key for a healthy life''.  
\textit{Content requests} differ since they concern the data or services provided in the app rather than the behaviour or qualities (i.e.~requirement) of the app. 
For instance, users might request to add soundtracks for an African music genre they miss in Spotify, or articles from a health magazine not yet available on the Newsfeed of Garmin Connect. 
In fact, user feedback can be even broader and not specific to the app itself. Users can comment the entire company or industry or  general emerging topics. 
For instance, Garmin Connect reviews might include general comments about Bluetooth or cellular network protocols (e.g., after new industry standards are developed). 
Spotify comments might concern the entire music and streaming industry or licensing models.

In addition, feedback often refers to existing app features. 
Users may simply describe how they use the app or certain features and why they find them useful (or not). 
Users might express \textit{improvement} ideas, for instance: 
``The lyrics feature is great. Sometimes the lyrics of my Arabic songs are out of sync. Please consider allowing users adjusting  the sync''. 
Or, ``I love the smooth integration into MyFitnessPal. 
Now, I can better monitor my calories stats particularly when on diet. 
And by the way, the dark theme is great too, particularly when reviewing my stats at night :)''. 
Such improvement requests and user experience reports help analysts and product owners gather additional details for a better understanding of the users and evolution of their products.

\begin{figure}[]
    \centering
    \includegraphics[width=0.9\columnwidth]{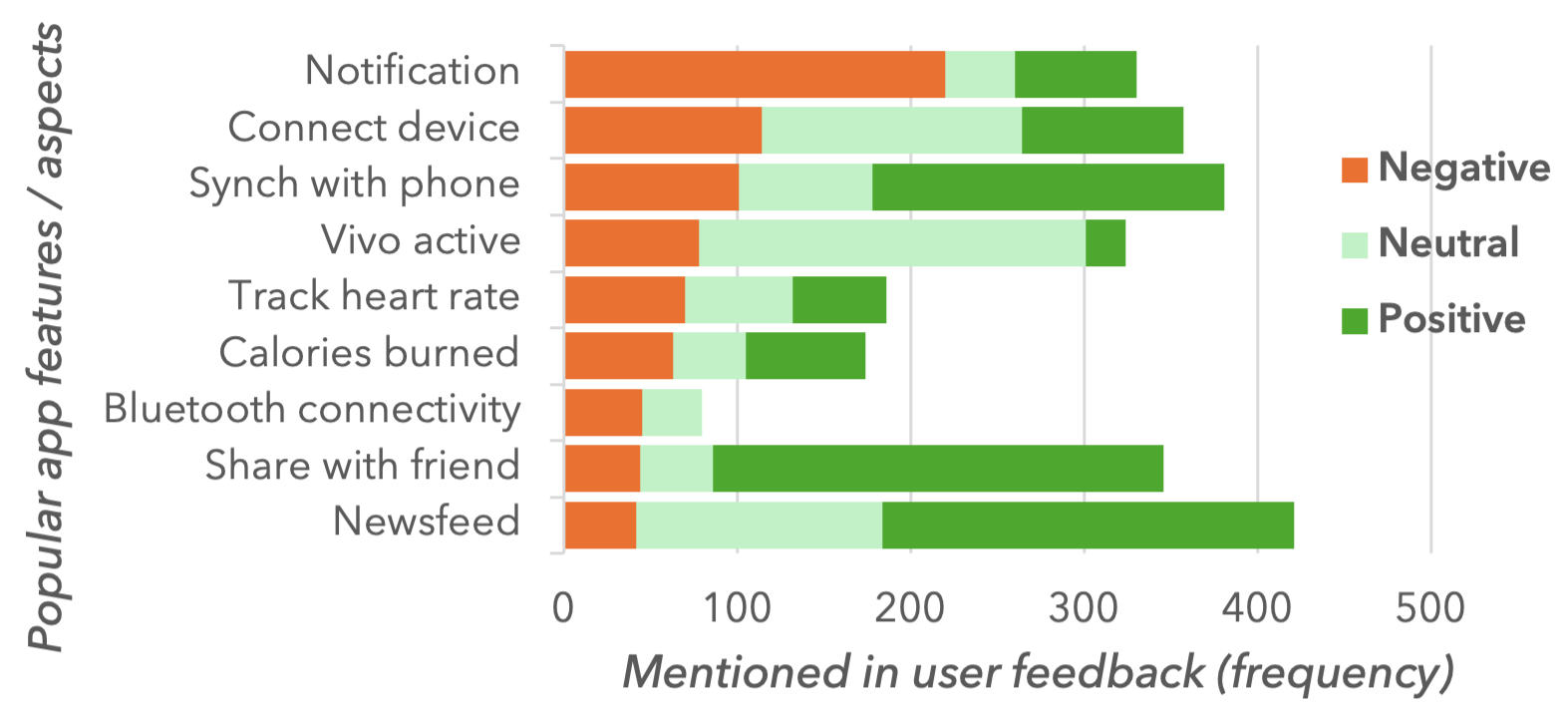}
    \caption{Examples of app features extracted  with the average sentiments.}
    \label{fig:popular-apps-sentiments}
\end{figure}

Monitoring how often features are being discussed with the average sentiments (whether users talk positively or negatively about the features) is a great way to summarise the satisfaction with implemented requirements, as Figure \ref{fig:popular-apps-sentiments} shows.
Software teams can monitor the sentiment trends over time, particularly after releasing new versions (i.e.~how is the sentiment towards features and topics changing over time). 
They can also compare features across the platforms (e.g.~why are iOS users more satisfied with the notification than Android users?).

\subsection{Monitoring and Improving Product Quality}
User feedback includes information that is useful for maintenance, testing, and quality assurance. 
Particularly, defects (or bugs) are often reported in app reviews and social media posts. 
For example, users might complain that the app crashed and all their data is lost. 
They might note that the screen was frozen when trying to accomplish a certain task. 
Or, that they accidentally clicked the purchase button but expected that the app would show an overview and ask for confirmation first.
Processing user feedback enables developers to gather such issue reports. 
A study by Haering et al.~suggests that about ~20\% of defects reported in app reviews might not (yet) be reflected in the issue tracker  \cite{Haering:ICSE:2021}. 
In the remaining feedback about bugs, developers can find additional contextual information about the bug, how it occurs. 
This helps reproduce and fix the bug. 
For instance: 
``I can’t open playlists shared via WhatsApp on my iPhone XR, iOS 12.1.4, Spotify 8.4.61''.
Even if users report redundant and uninformative issues like ``Can't open playlists'' or ``the app keeps crashing'' , the frequency of the reports informs about the bug severity.

Another useful aspect about feedback is that it simulates testing in the wild \cite{Gomez:Software:2017}, particularly when introducing experimental or immature features.
Reported issues might emerge in (user) contexts, which developers are unaware of or unable to create. 
For instance, a beta version might not be tested in a particular connectivity setting, on a special hardware, or with particular configurations (e.g.~sensors, versions, etc.). 
Furthermore, the quantity of feedback and the trends therein suggest where to focus the testing and quality assurance effort. 
If a certain quality requirement as performance  or availability is trending in the feedback channel, this is likely a sign of a high priority. 
Finally, testers can derive test scenarios and test data from feedback as users sometimes describe particular situations how they used the app. 


\subsection{Planning and Scoping}

One of the hardest tasks for developers,  managers, and product owners is to prioritise requirements and roadmap features or other development tasks. 
For instance, for an app like Garmin Connect, it is likely that there  will be more health-monitoring devices on the market, which can be connected with and ``integrated'' into the app, than available development resources. 
Prioritisation and scoping could go wrong if it is only based on  developer's assumptions (and gut feelings) or only on the opinions of a few stakeholders. 
Ignoring or hyping certain features as well as releasing them too early or to too late can have major consequences on the product success \cite{Martens:Software:19}. 
Collecting and processing ``enough'' feedback representing a large user base provides an overview about the user priorities---at least for those who submit feedback. 
Before taking prioritisation decisions, software teams can, e.g.~compare the average sentiments in the feedback with other data from workshops, market studies, and developers' own assessments.

Feedback can, for instance, be processed  to visualise the most frequently discussed app features as well as the average sentiments, as shown on Figure \ref{fig:popular-apps-sentiments}.
In this example, work on the \textit{notification component} should likely have a high priority, as it is mentioned frequently and often in a negative tone. 
Such analysis can also be split into certain geographical regions (e.g.~Central Asia vs.~North Europe), certain platforms (e.g.~Apple Watch vs. Samsung Watch), or certain types of users (e.g.~professional runners or music experts versus amateurs). 
In case of a gap between different user groups (e.g.~professional runners have a different opinion on showing a statistic than other users) the prioritisation might become a scoping issue. Product teams might, e.g., split the app into two different versions or decide to focus on one group. 

Most approaches presented in the literature describe data visualisation (analytics) solutions rather than fully automated or recommendation solutions. That is the goal should be to visualise the feedback in a way that it provides insights to decision makers rather than automate the prioritisation.
Nowadays, multiple commercial tools are also available to visualise the priority of users for entire markets. 
For instance, AppFlow\footnote{https://appfollow.io/} enables to compare the sentiment trends between different regions and topics. 
Tools like AppMagic monitor the sentiments about certain features over the entire market\footnote{https://appmagic.rocks/}.


\subsection{Supporting Users}
Since product reviews or social media posts are usually public and easily accessible, a feedback post by a certain user can be helpful for other users. 
Nowadays, most feedback channels enable users to like or rate the \textit{helpfulness} of a certain feedback item. 
Feedback can be helpful when it simply answers certain questions about the app not obvious from the documentation.
Other feedback posts might include detailed justifications, rationale, and decision criteria \cite{Kurtanovic:RE:2017}. These can be helpful to other users to decide whether to download, upgrade, or switch an app.
Large app vendors such as Spotify have entire support teams who engage with users in  social media to gather clarifications on the feedback, answer questions, or simply thank users for commenting and sharing their experience \cite{Martens:RE:2019}.

With the emergence of chatbots, the usefulness of feedback processing for support teams is raising to a different level. 
Bots can save support teams a significant amount of time to engage with users and collect needed information.
First, a bot can classify the feedback, e.g.~into a query or an issue report, to derive an appropriate reaction strategy.
A bot would also immediately engage with the user to collect missing details thus improving the feedback quality whenever needed \cite{Martens:RE:2019,Wolfinger:RE:22}. 
In case of an inquiry, e.g., on how to perform a certain task with the app or how to configure the system for enabling or disabling certain features,  the bot can try to answer directly or delegate to the support teams if not possible or not satisfactory. 
In case of a problem report or a suggestion, bots can check the issue trackers and potentially file unknown issues. 
Bots can also give users an immediate response to their reported bugs or requested features whenever reasonable, e.g. that the issue is known or that the suggested feature was discussed and explain why a difference decision was taken. 
This could dramatically improve the user experience and potentially their opinion.
 

\section{Feedback Processing Pipeline}
\begin{figure}[]
    \centering
    \includegraphics[width=0.95\columnwidth]{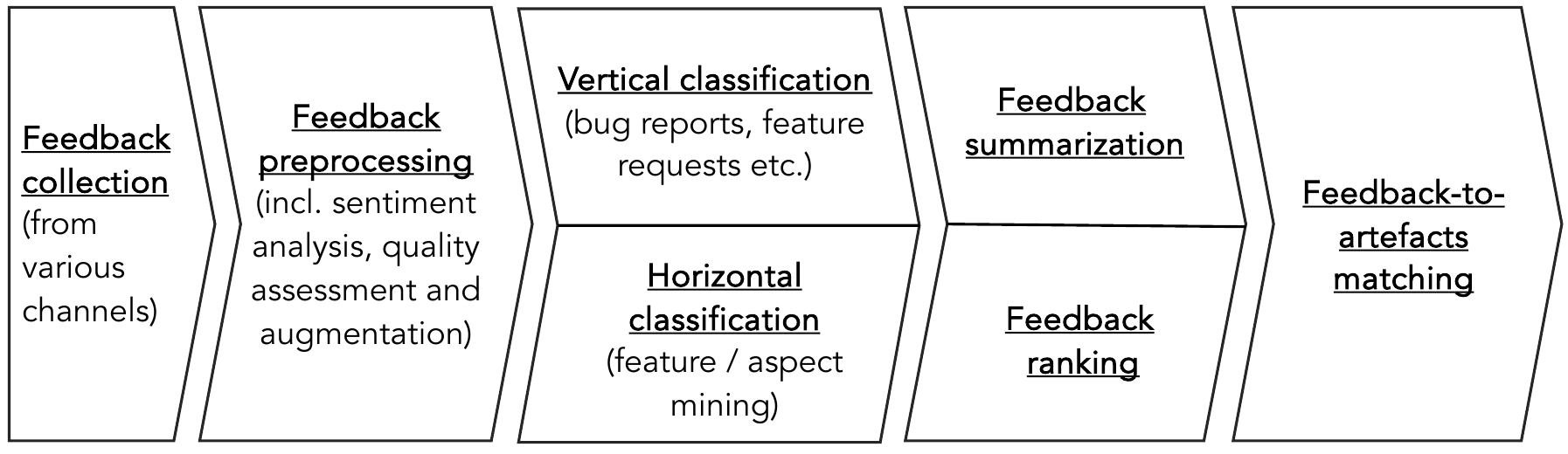}
    \caption{A general pipeline for feedback processing.}
    \label{fig:pipeline}
\end{figure}

The purpose of feedback processing is to extract useful and actionable information from a large amount of user feedback of varying quality.
Figure \ref{fig:pipeline} depicts a pipeline which starts by collecting the feedback from various sources as app stores and social media sites. 
Then, four major steps can be applied: 
1) feedback pre-processing (including quality assessment and augmentation), 
2) feedback classification (vertically along development tasks or horizontally along product features), 
3) feedback summarisation and ranking, as well as 
4) feedback-to-artefacts matching. 
We discuss each step, applicable NLP and Machine Learning techniques, as we as hints and pitfalls. 
In the accompanying material, there are  executable notebooks with examples for each step \cite{maalej2024z}.

\label{sec:pipeline}





\subsection{Feedback Preprocessing}
\label{sec:preprocessing}
Feedback preprocessing goes beyond common text preprocessing to A.~annotate it with sentiment scores or mask sensitive parts, B.~assess feedback quality and authenticity, C.~annotate feedback with demographic data to show how it represents certain groups, and D.~augment feedback with contextual data such as interaction traces. As these steps can be useful in the remaining pipeline, we recommended storing the preprocessed feedback in addition to the original data.

\subsubsection{A.~Text Preprocessing  and Sentiment Analysis} 

Common text preprocessing in NLP includes stemming or lemmatization, stop-word removal, or part-of-speech tagging. 
Depending on the ML technique used in the following steps, certain textual preprocessing might be needed. 
For instance, reducing the inflection forms of the terms ``runners, running, runs, ran'' into ``run'' might help retrieve and cluster feedback items around this topic. 
However, as user feedback tend to be short, text preprocessing can also be counterproductive, leading to information loss. 
Often, tokens such as the word ``not'' or the postfix ``ed'' (to denote past tense) are needed to understand the context, or train a classifier. 
Thus, text preprocessing should be used with care on  feedback, particularly when using large models like BERT or GPT that are sensitive to word context. 

\textbf{Masking} is a common feedback preprocessing step. 
The goal is to replace specific word tokens, such as hashtags, $@$-mentions, links, screenshots, usernames or personal information, app versions, app names, vendor names, or other specific words with a placeholder like $<$hashtag$>$, $<$mention$>$, etc. 
These token can then be used as machine learning features to train feedback classifiers. 
Moreover, the masking is often required to meet certain privacy requirements when using and re-sharing the feedback of users, e.g.~in ML models and visualisations.

Moreover, users will likely spend little time writing their voluntary feedback. 
Often, this happens on a mobile phone without paying much attention to language \textbf{consistency and spelling}.
Therefore, lower casing, spell and grammar checking and autocorrection techniques can be used to harmonise feedback (e.g.~the same word spelled differently as ``Lyric'', ``lyrics'', and ''lirics'').
Common libraries, such as \textit{textdistance} for Python\footnote{\url{https://pypi.org/project/textdistance/}}, use edit distance metrics such as Levenshtein distance to calculate the distance between the user terms and reference words from a dictionary.
These libraries find word permutations within a pre-defined distance and suggest fixes, as well as spelling errors such as scrambled letters, missing, or redundant letters.

\textbf{Translation} is another useful preprocessing step. 
Research had a strong focus on English language so far.
For software with an international user base or for non-Anglophone countries, feedback processing in different languages can be challenging, e.g.~due to missing labelled datasets. 
One possible workaround is to translate feedback into English language using tools such as DeepL\footnote{\url{https://www.deepl.com}}, or pre-trained models from research~\cite{fanEnglishcentricMultilingualMachine2021}.
However, not all languages are always available\footnote{\url{https://support.deepl.com/hc/en-us/articles/360019925219-Languages-included-in-DeepL-Pro}} and the quality of translation differs significantly between languages~\cite{fanEnglishcentricMultilingualMachine2021}. 
Another possibility is to train separate ML models for each language, which might require labelled datasets from all involved languages and is therefore more demanding.
The feasibility of such approaches has been demonstrated by Stanik et al.~for feedback classification using traditional machine learning and deep learning \cite{stanikClassifyingMultilingualUser2019}.
Their dataset includes feedback in German and Italian. 

Finally, \textbf{sentiment analysis} is a common and important feedback preprocessing step \cite{Guzman2014}. 
Its goal is to extract the sentiment polarity of a given text. For instance, the phrase ``excellent visualisation'' is rated positive, ``the track download is really annoying'' is rated negative, and ''I downloaded this app last month'' rather neutral. 
The output is usually a polarity score between -5 (very negative) and +5 (very positive). Often the scale can be customised (e.g. -3 to +3) or simplified to positive or negative. 
Using a differentiated scale (at least including a neutral category) avoids conflating the analysis results. 
The annotation with the scores can be fine-grained at the term level (might help understand reviews), at the sentence/phrase level (recommended for a differentiated processing as a review might include multiple sentiments), or coarse grained at the review/entire feedback level (help retrieving extreme reviews). 

There are two main sentiment analysis approaches. 
Lexical sentiment analysis is based on a dictionary of terms with their sentiment polarities. 
Semantic approaches learn combinations of terms and the whole sentence context. 
Popular tools include Lexalytics, TextBlob, or SentiStrength. 
The accuracy of these tools is usually high, reaching a human level assessment. 
However, negations, exaggerations, jokes, or sarcasm remain challenging for automated analysis.
Large language models can also be easily prompted to estimate the sentiment of a text (see Example \ref{prompt:aspects}), usually with a similar accuracy to state of the art.




\subsubsection{B. Quality Assessment}
An often ignored preprocessing step is to check the feedback for different quality criteria such as authenticity, ambiguity, or redundancy.
Feedback can be marked with quality scores or ultimately filtered out. 

Martens and Maalej showed that \textbf{unauthentic reviews} (aka fake or incentivised, non-spontaneous and thus possibly wrong and misleading) exist to a significant extent even in moderated channels such as app stores~\cite{Martens:EMSE:2019}. 
To emphasise the importance of detecting fake reviews, they showed that many fake reviews are classified as ``regular'' bug reports and feature requests by state-of-the-art classifiers.
The authors built a dataset of fake app reviews. They showed that an accurate classification should also be based on user data rather than only on the feedback text itself, since the textual properties do not significantly differ from authentic reviews~\cite{Martens:EMSE:2019,mukherjeeWhatYelpFake2021}. 
The problem with fake unauthentic review is that it can be completely misleading as a basis for decision-making in requirements engineering. 
Therefore, this preprocessing step might be particularly important if not much redundancy is available in the entire feedback dataset.

Moreover, \textbf{language ambiguity} can impede the ability to analyse and use feedback.
For example, referring to a previously mentioned aspects or app features  as ``it'' can result in incomprehensible review or a mismatch in processing.
The problem of language ambiguity, including the use of passive voice or compound nouns, is not exclusive to user feedback and has been extensively studied in requirements engineering \cite{Montgomery:REJ:22}.
However, it might be particularly common in feedback as users might not be native speaker, not trained for the task or simply do not have enough task and incentive to write a precise readable text.
Also automatically understanding sarcasm, irony, ambiguity, and informal language also remain mainly unsolved ~\cite{khuranaNaturalLanguageProcessing2023}.
Sarcasm detection is particularly relevant for sentiment analysis of user feedback, since it can make the difference between praise and a problem report~\cite{vermaTechniquesSarcasmDetection2021}.

Another form of low quality feedback is the usage of toxic, abusive, or \textbf{offensive language}.
Insults are sometimes directed towards the developers, such as  ``entire department needs to be fired'' or ``developers need shooting''.
The detection of offensive language in user feedback is not trivial, as the same words can have different meanings in different contexts (e.g.~``kill the background process'' vs.~``developers should kill themselves'').
Multiple approaches to detect offensive speech have been proposed in the social media mining literature~\cite{davidsonAutomatedHateSpeech2017,vanakenChallengesToxicComment2018,weiOffensiveLanguageHate2021}.
Finally, feedback might be too short, uninformative, out of scope, or repetitive.

\subsubsection{C.~Annotation with Demographics}
Feature-rich software is often built for a wide variety of use cases and users around the world.
This gets reflected in the diversity of feedback and submitting users.
Demographic attributes like primary language, experience, culture, communication channels, and even personality traits~\cite{Biryuk:AffectRE:23} might have an impact on the feedback content and linguistic properties as recent research shows~\cite{guzmanUserFeedbackApp2018,Fischer2021}. 
Therefore, whenever possible, such metadata about submitting users should be collected or extracted from the text.

Tizard et al.~\cite{Tizard2021} surveyed software users about their feedback habits, software usage, and demographic information.
Their results show that only a subset of users provides feedback.
Those users tend to have different demographic properties than the general population.  
This should be carefully considered when processing feedback.
The authors found differences in the amount and type of written feedback with respect to gender, age, and professional affiliation.

The motivation to provide feedback (e.g.~to show appreciation or get help) also tends to depend on the communication channel, e.g.~social media vs. product forums.
This is an important factor in the selection of the feedback source with regard to content, but also technical feasibility.
For example, if feedback channels have different technical restrictions regarding text length, it has to be considered when using NLP techniques and training feedback classifiers.
Sampling from the ``right'' communication channels is also important, since the distribution of motivations and topics differ between them.
In an ideal scenario, demographic information of the users would be available to the software teams and could be used to create a more diverse feedback sample.

For cross-platform software, feedback sampling has to encompass users from all platforms.
For example, sampling feedback of an app that has a web browser and an Android version has to be adjusted accordingly. 
Selecting the feedback source is an important step at the beginning of the sampling process.
Practitioners usually have access to support tickets, emails, distribution/review platforms, social media, and product forums~\cite{vanoordtRoleUserFeedback2021}.
If multiple channels are available, different sources can be aggregated by semantically matching similar feedback~\cite{oehriSameSameDifferent2020}.

Fischer and Guzman also showed that users from cultures that are more collectivistic and uncertainty avoidant write longer feedback texts~\cite{Fischer2021}.
This should be considered when choosing the appropriate NLP and classification algorithms but also for sentiment analysis and the interpretation part.
Recently, Biryuk and Maalej~\cite{Biryuk:AffectRE:23} also argued that personality traits and cognitive aspects might also impact the feedback quality and style and should, if possible, be considered too.

\subsubsection{D.~Feedback Augmentation}
\label{ssec: Feedback Augmentation}
A large portion of textual feedback is written by non-technical users who do not necessarily understand the information needs of requirements analysts and developers.
Especially technical details such as device settings, app and system versions, crash logs, or interaction data might be unclear to users and thus not shared in their feedback \cite{Martens:RE:19}.
However, these details are often crucial for understanding the feedback and reproducing or isolating reported issues.
If applicable and acceptable, feedback processing approaches should consider  augmenting the text with additional technical and contextual data in form of traces and logs collected as implicit feedback \cite{maalejDataDrivenRequirementsEngineering2016,Gomez:Software:2017}.

Implicit feedback requires observing users during the software usage and collecting background data without explicit user actions.
Such data can be interaction data (e.g.~click sequence), crash reports, hardware information, or sensor data (e.g.~light data to understand feedback about dark mode of sound lyrics feature).
One hard challenge is to specify beforehand what data should augment feedback, as some implicit feedback might be more valuable to analysts than others.
For example, crash reports are particularly useful when the users are complaining about an app crash in an app review \cite{Gomez:Software:2017}.
Libraries as Crashlytics, Sentry, or Adobe Analytics provide powerful features for collecting and analysing implicit feedback.
This can then be attached to low-quality brief feedback with the consent of users, or correlated with explicit feedback in general \cite{maalejDataDrivenRequirementsEngineering2016}.

Another useful implicit feedback is the specific app features or component used and to which written feedback refers. 
Stanik et al.~describe a method to learn app feature usage from interaction data~\cite{stanikWhichAppFeatures2020}.
They leverage the Android accessibility framework to gather interaction events such as clicks, notifications, edits, selections, scrolling, view focus, etc.
They subsequently use a ML model to classify patterns in the interaction event data into feature usage patterns.
Combining feature usage patterns with written feedback can improve the understanding of the usage context.
For example, a very negatively written review can lead to the conclusion that there is a major problem with the app.
However, if the review concerns only a rarely used, wrongly used, or never used feature, the impact of the problem can be different. 

\subsection{Feedback Classification}
The goal of the preprocessing steps is to transform the original text into a ``better/usable text'' and annotate it with additional metadata like sentiment scores and demographics. 
The resulting text can thus be processed more reliably and leading to more insights in the core feedback processing steps.
One of these steps is feedback classification. We distinguish between vertical classification (along the targeted development activities) and horizontal classification (along the commented-ed app features or other app-specific aspects).

\begin{figure}[]
    \centering
    \includegraphics[width=0.8\columnwidth]{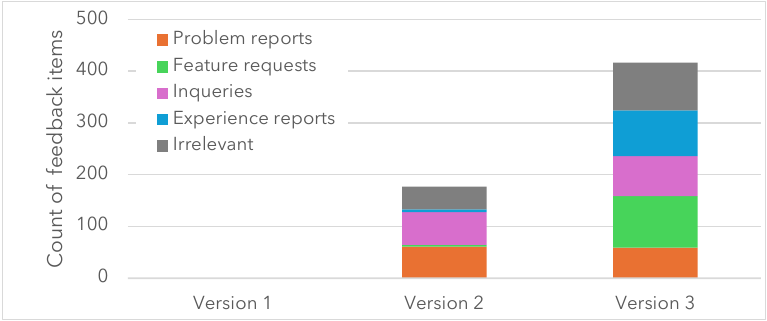}
    \caption{Vertical feedback classification visualised across multiple app versions.}
    \label{fig:classification}
\end{figure}

\subsubsection{Vertical Classification}
Feedback items can be automatically classified into predefined categories, most notably into: bug reports, feature requests, user experience reports, enquiries, and uninformative/irrelevant \cite{maalejAutomaticClassificationApp2016}. 
Vertical classification is particularly useful for the feedback actionability, as it enables directing the feedback into different organisational units or teams. 
For instance, development and quality assurance teams are more interested in bug reports; analysts and product owners in feature requests and feature ratings, UX and documentation teams in user experience reports, support teams in inquiries. 
Moreover, classification enables overviewing and comparing the amount of feedback for different categories, as shown on Figure \ref{fig:classification}.

Most feedback classification approaches train a classifier using the textual features based on a labelled dataset \cite{maalejAutomaticClassificationApp2016}. 
However, other numeric or ordinal features such as the feedback rating, length, tense, sentiment scores, or demographics can be used too and might lead to more accurate classification. 
If teams have no access to datasets or machine learning techniques, even simple approaches using regular expressions and particular keywords like ``bug, problem, issue, crash, feature'' can lead to  satisfactory results \cite{maalejAutomaticClassificationApp2016} particularly for bug reports.

The target classes are independent of the app itself, which makes it possible to use datasets from other apps. 
Common classes include bug report, feature request, inquiry, relevant/irrelevant, positive/negative feedback, user experience description, as well as specific non-functional requirements such as security, usability, and performance.
The main challenge for vertical feedback classification is to obtain a high-quality labelled datasets, which are meanwhile frequently available in literature, although often not for commercial use. 
In contrast, the biggest advantage is that classifiers usually work well independently of the app. 
The classifier can be trained on feedback data from one app and used on another.

Modern feedback classification approaches are based on transformers, such as BERT~\cite{DBLP:conf/re/MekalaIGPL21}, S-BERT~\cite{10263218}, CamemBERT~\cite{Wei2022} or RoBERTa~\cite{Araujo2020FromBT}, which learn the context of words through embeddings. 
Word Embeddings are representations of words as multidimensional vectors, whereas the distance indicates the semantic similarity of the words. Various algorithms for word embedding creation exist including word2vec, GloVe, BERT, and ELMo.
More recently, and with the emergence of Large Language Models, it became also possible to classify feedback using zero-shot or few-shots learning with accurate results. An example of such a prompt is shown on Example \ref{prompt:classify}.
Wei et al.~\cite{Wei2023} showed that GPT-3.5 achieved a top performance in the vertical classification, with an F1 score exceeding 0.85 in the classification of 6000 English reviews into three distinct categories.

\begin{tcolorbox}[arc=0mm,width=\columnwidth,
                 top=1mm,left=1mm,  right=1mm, bottom=1mm,
                 boxrule=1pt] 
\begin{example} 
\label{prompt:classify} 
Please classify the following app reviews into problem report, feature request, user experience description, or irrelevant. Be concise. \\
\`{} \`{} \`{} 
 \{reviews\}
\`{} \`{} \`{} 
\\ A review can belong to multiple classes. 
In the least possible amount of words, explain why the classification decision was made. Refer to the text.  
\end{example}
\end{tcolorbox}

It is important to note that one review might include more than one class. 
For instance, users might describe their experience with a certain feature and request another feature in one review. 

\subsubsection{Horizontal Classification (and Clustering)}
The vertical classification is straight as the classes are independent from the app at hand and easy to specify beforehand. 
In the horizontal classification, the classes refer to particular features, components, or topics of the app:  such as ``notification'' and ``sign in'' or ``device connectivity'' and ``lyrics''. 
The main challenge here is that these classes are not known beforehand (depend on what users are discussing), might change over time, and are hard to create as they often concern specific apps for a certain time. 
Therefore, clustering is often needed for this step.
Feedback clustering refers to grouping feedback entries together based on their semantic similarity, usually as they concern the same or similar topics. 
The feedback in the same cluster are more similar to each other than to those in other clusters \cite{Stanik2021}.

As user feedback are in textual format, applying common clustering algorithms is not straight.
String similarity would quickly reach its limit as the text is usually still heterogeneous, informal, and usually includes multiple sentences (longer than a few words).
Different methods were suggested to convert feedback into numerical vectors, including frequency-based approaches (e.g.~Bag of Words, N-Grams, TF-IDF) and embedding-based approaches (e.g. Word2Vec~\cite{Mikolov2013}, Universal Sentence Encoder~\cite{Cer2018}, BERT~\cite{Devlin2019}).
Wei et al.~\cite{Wei2023} recently found that text representation methods have a crucial impact on the clustering performance, and neural network-based approaches outperformed frequency-based approaches, which is unsurprising as they tend to represent richer context of the text and thus its meaning  \cite{Haering:ICSE:2021}.
When feedback entries exhibit similarity in meaning, their respective numerical vector representations tend to exhibit proximity. 
Numerical vectors created from feedback entries are of high dimensions and may suffer from data sparsity and high computation costs.
To overcome these limitations, dimension reduction techniques as UMA~\cite{2018arXivUMAP} can be applied to reduce the data dimensionality while keeping meaningful information of the original data \cite{Stanik2021}.

Clustering algorithms then group these numerical representations of feedback based on their distance, i.e.~their semantic similarity.
Popular clustering algorithms includes partitioning clustering (e.g.~K-Means), dense-based clustering (e.g. DBSCAN~\cite{Ester1996}, HDBSCAN\cite{McInnes2017}), hierarchical clustering, graph clustering (e.g. Chinese whispers~\cite{Biemann2020}), topic modelling (e.g. LDA~\cite{Campbell2015}), etc.

Researchers have evaluated these approaches on user feedback with mixed results. 
Chen et al.~\cite{Chen2014} employed LDA to assign topics to each user review.
Guzman and Maalej \cite{Guzman2014} used LDA on term collocations extracted from app reviews to identify coarse grained app features. 
Scalabrino et al.~\cite{Scalabrino2019} applied DBSCAN to identify groups of related reviews.
Recently, researchers \cite{Stanik2021,Devine2022,Wei2023} used HDBSCAN for the discovering of topics from user feedback, where the work of Stanik et al.~\cite{Stanik2021} shows that HDBSCAN can result in partly cohesive and meaningful clusters.
Wang et al.~\cite{Wang2022} extract problematic features from user reviews and perform Chinese Whispers, K-means, and LDA to cluster these features. They found that Chinese Whispers outperforms the two other clustering methods in precision, recall, and F1 score. 

Generally, the results are of mixed quality as some clusters are cohesive and meaningful, e.g.~referring to ``sharing playlists'' or ``Bluetooth connection'' while other topics are not.  
Therefore, this processing step should be done iteratively with an analyst-in-the-loop. 
Recurrent, popular topics can also be specified as particular classes to explicitly learn and predict.
It is also important to manually review a small sample of the original feedback to understand and choose a precise name of the cluster.
If the entire list of features is available (e.g.~from the app description page or from the release notes), this process might be re-framed into a classification problem \cite{johannSAFESimpleApproach2017}.  
Finally, LLMs can be prompted as well to retrieve popular topics in the reviews, e.g. as shown in Example \ref{prompt:aspects}. 

\begin{tcolorbox}[arc=0mm,width=\columnwidth,
                 top=1mm,left=1mm,  right=1mm, bottom=1mm,
                 boxrule=1pt] 
\begin{example} 
\label{prompt:aspects} 
Your task is to perform aspect-based sentiment analysis on app reviews. Please extract app features and sentiment polarities (positive, neutral, negative) from reviews: \\ 
\`{} \`{} \`{} 
 \{reviews\}
\`{} \`{} \`{} 

The output should be a list as JSON format like this: \\
\[\{\{ {}
 "feature": feature,
 "polarity": polarity
\}\}\] {}

\end{example}
\end{tcolorbox}


\subsection{Feedback Summarisation and Ranking}
While it is useful to classify and cluster feedback to get an overview about the frequency and sentiment for each category, a class or a clusters usually contain numerous feedback items. 
It is thus often impractical for developers to read all of them to understand the content.
To address this issue, summarisation techniques can be applied to create a succinct summary for a group of feedback. 
For instance, the same bug is often reported numerous times in app reviews, partly with redundant information, partly with additional context. 
If accurately done, summarisation can lead to one single issue report with the various noteworthy details. 
Alternatively, feedback ranking orders the list of feedback items showing the most relevant feedback items first.

\textbf{Summarisation techniques} can be broadly classified into two categories: extractive summarisation and abstractive summarisation~\cite{DBLP:books/sp/MehtaM19}.

\textit{Extractive summarisation}
involves selecting the most important sentences or phrases from the source text and reorganising them to form a summary. 
Topic models such as LDA are widely used for app reviews summarization \cite{Chen2014,Guzman2014}. 
LDA is a probabilistic distribution algorithm which uses Gibbs sampling to assign topics to documents. 
A topic is a probabilistic distribution over words.
Each feedback item can be associated with different topics, and those topics are associated with different words with a certain probability.
Devine et al.~\cite{Devine2022} select the most representative unigrams, bigrams, trigrams, or sentences of a cluster by calculating their semantic similarity. 
They evaluated the effectiveness of these four methods on understanding the meaning of a feedback cluster and found that the sentences can better summarise important requirement relevant information from a cluster.

\textit{Abstractive summarisation}
creates a summary that captures the core ideas of the source text. 
The generated abstractive summaries may contain new phrases or sentences not explicitly present in the source text.
Hark et al.~\cite{Harkous2022} introduced a hierarchical summarisation approach to generate a short phrase as the summary of a user feedback cluster.
They first apply T5 \cite{Raffel2020} to generate the issue summarising the main topics of a user feedback, then cluster these issues, and finally, use T5 to generate a theme for the issues of the same cluster.
Wei et al.~\cite{Wei2023} applied LLMs, including ChatGPT and Guanaco \cite{Dettmers2023}, to generate sentences for a cluster of user feedback.
Their empirical evaluation shows that abstractive summaries generated by LLMs have a higher quality than extracting most representative sentences.
A sample prompt is shown Example \ref{prompt:summarize}.

\begin{tcolorbox}[arc=0mm,width=\columnwidth,
                 top=1mm,left=1mm,  right=1mm, bottom=1mm,
                 boxrule=1pt] 
\begin{example} 
\label{prompt:summarize} 
Please summarise all the following app reviews into one sentence:
\`{} \`{} \`{} 
 \{reviews\}
\`{} \`{} \`{} 
\end{example}
\end{tcolorbox}

\subsubsection{Feedback Ranking}
When analysing a large number of feedback, ranking methods help  prioritise a) the entire clusters (i.e.~for requirements prioritisation and scoping or to determine areas that require immediate attention) as well as the single feedback items inside each cluster.

Multiple factors can be considered when ranking review clusters \cite{Chen2014,Wei2023}:
\begin{itemize}
\item Number of reviews.
Clusters with a higher number of reviews should be given greater priority, as they likely represent topics that are of concern to a larger portion of users.

\item Average rating and sentiment.
Clusters with lower average ratings (or average sentiments) should be prioritised higher, as they may suggest higher levels of dissatisfaction among users regarding specific aspects of the application.

\item Helpfulness score / Number of ``thumbs up''.
On platforms like Google Play, users have the option to indicate helpful reviews by clicking the ``thumbs up'' button. 
The number of ``thumbs up'' received by a review can indicate its importance, as reviews that are liked by more users should be given higher priority in the ranking process.

\item Time Series Pattern.
Each app review contains a timestamp.
By plotting the number of reviews within distinct time windows, we  obtain the trend over time.
As shown on Figure \ref{fig:time-series-pattern}, the horizontal axis represents the time windows, while the vertical axis $p(g,k)$ signifies the quantity of reviews belonging to a particular group $g$ in a given time frame $k$.
The pattern in the left part of the Figure illustrates a sharp increase within a specific time frame, which could be attributed to the a new bug or request due to an update e.g. in the operating system. 
Conversely, the right pattern in the Figure depicts a swift decline, indicating an alleviation or resolution of an old bug or request.
A cluster with a time series pattern similar to the left side should have a higher priority than those of the right side.

\begin{figure}[]
    \centering
    \includegraphics[width=0.6\columnwidth]{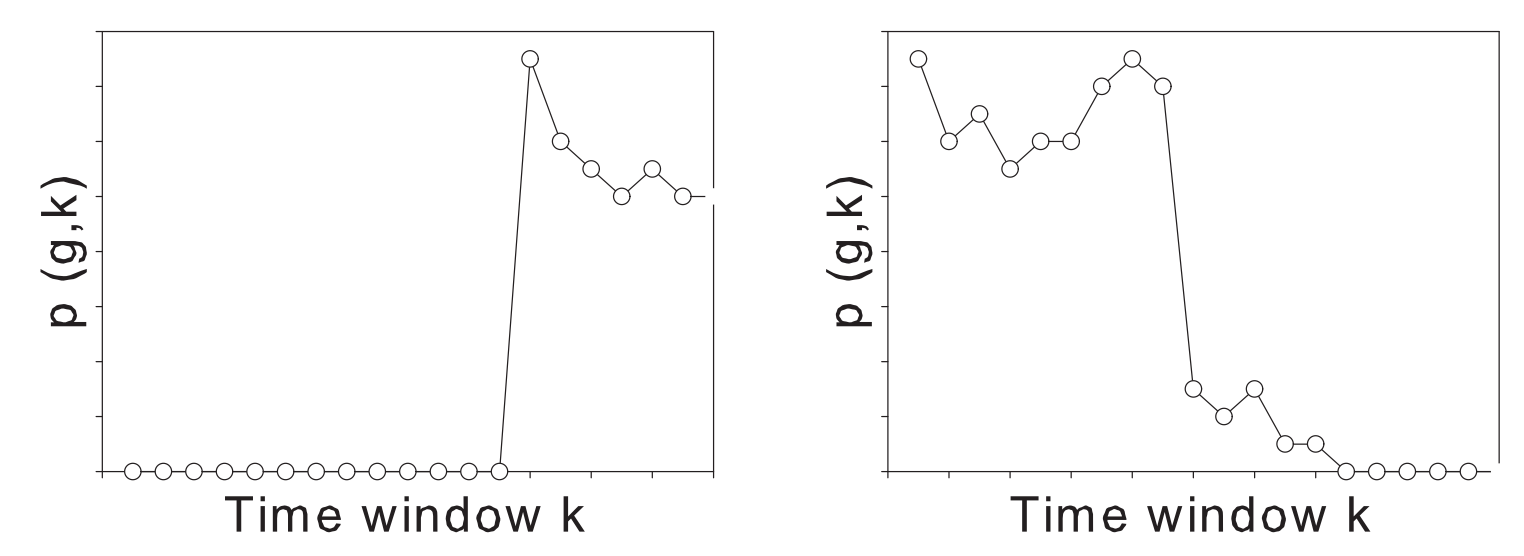}
    \caption{Representative Time Series Patterns (adapted from \cite{Chen2014}).}
    \label{fig:time-series-pattern}
\end{figure}
\end{itemize}
In practice, feedback ranking should incorporate different factors, where users can, e.g. select the main factor or assign varying weights to each factor.


\subsection{Matching Feedback to Development Artefacts}\label{sec:integration}
Preprocessed, possibly filtered, classified, clustered, and eventually summarised or ranked feedback should be brought to the attention of stakeholders, ideally in their work environment fitting their workflows. 
Researchers started recently looking at this challenge and developing tools and approaches to match and integrate processed user feedback into the engineering workflows.

Most notably, \textbf{project issues} steered and managed in issue trackers (like Jira\footnote{https://www.atlassian.com/software/jira}, Redmine\footnote{https://www.redmine.org/}) can be matched to user feedback. 
Issues are assigned to particular project members and certain releases.
They get discussed, resolved, tested, and closed. 
Issue trackers are very popular among agile teams to manage work items including user stories and epics. 
However, there might be a gap between what happens in the issue tracking system and what happens online in the user space.
To address this gap, approaches like DeepMatcher~\cite{Haering:ICSE:2021} matches bug reports in issue tracking system with related user reviews from app stores.
First, this helps developers track whether an issue reported in an app review is already filed as a report in the issue tracker.
Moreover, this enables collecting additional contextual information about the bug such as unique contextual information reported by certain users.

Unlike feedback from social media and product forums, items in issue trackers have more structured fields and stricter reporting and management rules.
For example, Jira allows issue reporters to use bullet points, highlight source code snippets (e.g. \{code:java\} ... \{code\}) or include screenshots \footnote{\url{https://jira.atlassian.com/secure/WikiRendererHelpAction.jspa?section=all}}. 
This additional information might simplify downstream NLP tasks as they might denote very specific context.
User feedback on the other hand rarely includes such structures.

In practice, user feedback from product forums can be used as a source to create new issue tracker entries or matched with existing entries to complement potentially incomplete information~\cite{Haering:ICSE:2021}.
Haering et al.~and Tizard et al.~have demonstrated, that it is possible to match app reviews and product forum posts to issue tracker entries using deep learning~\cite{Haering:ICSE:2021,tizardSoftwareRequirementsEcosystem2023}. 
First two embeddings need to be created: one for issues in the issue tracker and one for user feedback. 
Then, the similarity of the embeddings is compared and feedback items that pass a certain threshold are matched with corresponding issues.

Other artefacts documenting the \textbf{system features} and their relationships can also be matched to feedback. 
This particularly includes formal artefacts like goal models, or rather informal artefacts like app pages on app store \cite{johannSAFESimpleApproach2017}, blogs, and release notes listing app features from the vendor's perspective. 
Goal models are used in requirements engineering research and to some extent in practice too. 
They refine requirements by asking ``why'' and ``how'' questions. 
The creation of a goal model enables requirements engineers to describe, reason, and visualise software features and their relationships.
Given that a Goal Model should represent all features of a software, it is desirable to combine the reviews with the nodes of the goal model automatically, thereby offering a visualisation of user satisfaction levels concerning specific features and identifying those that necessitate improvement~\cite{Liu2020} or are missing.
By extracting feature requests from user reviews and comparing these with the existent goal model, it is possible to automatically suggest potential features for developers~\cite{Gao2020}.


Finally, linking user feedback with \textbf{source code} can help developers understand which code section should be updated to satisfy the user's need.
For instance, Palomba et al.~\cite{Palomba2017} proposed CHANGEADVISOR, a tool that extracts user feedback relevant from a maintenance perspective, clusters it according to similar user needs, and determines the code artefacts that need to be maintained to accommodate the requested changes.

\section{Data Quality for Feedback Processing and Management}\label{sec:quality}

Data-centred decisions can only be made if the necessary data is available and reliable.
Thus, data quality (short DQ) management~\cite{LeePFW2006,FanG12,BatiniS16} 
plays an important role in decision support tools in general, and feedback processing in particular. 
Surprisingly, quality management for user feedback  has received rather a little attention so far. 
Since user feedback is data, 
many concepts from traditional data management 
can presumably be adopted.
Although user feedback itself is often not structured in attribute-value pairs, it can be accompanied by metadata (e.g. timestamps, demographics or implicit feedback as described in Section~\ref{sec:preprocessing}).

In this Section, we provide a brief overview of four fundamental areas of \DQStart management,  which has a main focus on relational, i.e. structured, data. 
We discuss how these areas can be potentially used for feedback processing.
The first area deals with \textit{what} data quality means and how it is characterised.  
The second area deals with how the various aspects of \DQ can be \textit{quantified} and efficiently calculated. 
The third area deals with maintaining a (good) quality for datasets over time and avoiding degradation due to errors and misuse. 
The fourth and final area deals with improving a  (bad) quality for a dataset. 
This includes the removal of errors, duplicates, and inconsistencies.

An important concept in \DQ management is  \textbf{fitness for use}.
That is, data that is good enough for one application may be insufficient for another.
Therefore, a universal definition and treatment of \DQ is neither reasonable nor desired, 
making appropriate \DQ management even more difficult. 

\subsection{Data Quality Semantics}
\label{sec:DQ_Semantics}

Mistakenly, \DQ is often reduced to the correctness (or accuracy) of data~\cite{WangS96}. 
However, \DQ extends far beyond this, including, among others: completeness, timeliness, level of detail, uniform presentation, or the relevance of the data. 
For example, submitting negative feedback on the connectivity of ``smartwatches'' might be correct, but would not correspond to the required level of details as many brands and models of smartwatches are supported by the app and distinguishing between them is necessary for understanding the feedback.
In addition, in the case of user feedback, trustworthiness is a particularly important quality factor as users may intentionally provide false information~\cite{Martens:EMSE:2019}.

For this reason, \DQ is usually divided into several dimensions, each of which is intended to capture a different aspect of quality. 
Since \DQ is difficult to grasp and there are often (slight) differences in its understanding, no definition of \DQ dimensions has yet become established. 
However, many well-known definitions have been proposed over the last three decades.
They are typically based on theoretical models (e.g.,~\cite{WangS96}),
empirical studies among data consumers (e.g.,~\cite{WandW96}), or human intuition (e.g.,~\cite{Redman1996}).
Recent efforts for defining \DQ dimensions have been made by Mohammed et al.~\cite{MohammedHNS24}, the ISO 8000 International Standard for Data Quality, and the ISO/IEC DIS 5259-1 Standard for Data Quality for Analytics and Machine Learning.
However, these approaches primarily refer to relational data while other dimensions might be relevant in the case of user feedback.

\subsection{Data Quality Measurement}
\label{sec:DQ_Measurement}

Conceptually defining what \DQ means is one thing. Measuring it quantitatively is another.  
The first challenge is to capture \DQ semantics properly.
The second is to measure it with the available information, which may vary between use cases.
The third challenge is to measure with an acceptable computational overhead.

The first challenge is particularly difficult
because, in most cases, we do not want a binary score (e.g. the feedback is correct or incorrect), 
but a graded score that allows a comparison between several imperfect datasets. 
For instance, we might aim to give a quality score in the range between 0 (very bad) and 1 (very good) for each feedback item or for a set of feedback items.
It is often not easy to see which dataset is of better quality. For instance, it remains unclear 
what is worse: a dataset with one significant data error (e.g.~a false defect which is actually a usage mistake) or a dataset with two minor data errors (e.g.~two imprecisely reported defects).
Defining a (partial) order can thus be toilsome but  necessary for a graded evaluation.

In addition to the order of the calculated quality scores, their distances should also reflect the actual distances in the quality of the corresponding datasets.  
For instance, the score should not only satisfy an ordinal scale but rather an interval or ratio scale.
To address these issues, Heinrich et al.~\cite{HeinrichHKSS18} proposed a list of properties which should be met by well-designed \DQ metrics.

While some \DQ dimensions can be calculated on the available data, others require additional information, 
such as knowledge bases (e.g.~about the domain or to determine semantic relationships) 
or reference datasets which act as ground truth. Reference datasets can either be obtained externally 
(e.g.~hardware specifications from vendors or legal requirements) or created manually by experts for a small subset of the data. 
Accuracy~\cite{BatiniS16} is an example \DQ dimension that requires ground truth is, 
because it requires comparing with the correct data value. 
Reference datasets are often hard to obtain, particularly in opinionated data like feedback. 
Approximations might thus present a useful alternative. 

For accuracy, a distinction is often made between syntactic and semantic errors~\cite{BatiniS16}. 
While syntactic errors are characterised by a violation of the data type or domain at hand and thus can be determined without any ground truth 
(e.g. feedback referring to a non-existing version of an app), 
semantic errors do not contradict these domains and thus cannot be identified without the ground truth
(e.g.~feedback referring to an existing but not the actually used version of an app). 
Methods from data profiling (see Section~\ref{sec:DQ_Improvement}) discover semantic rules within the data
(e.g.~rules that state which app features were available from which app version).
Such rules may help convert some types of semantic errors into syntactic ones to identify them even without a ground truth.
To address the last challenge, \DQ metrics should be calculated efficiently in the context of Big Data 
and incrementally in the context of dynamic data~\cite{SchelterLSCBG18}.
Ehrlinger and Wöß overview existing tools for measuring and monitoring \DQ \cite{EhrlingerW22}.

\subsection{Data Governance \& Data Quality Assurance}
\label{sec:DQ_Governance}

Ensuring high data quality requires measures along the entire data pipeline. 
This starts within the companies or organisations that collect and maintain the data~\cite{Eryurek21}, as the app vendors or the app store providers. 
Data governance includes internal company standards and guidelines that affect data collection, processing, and insertion. 
For instance, a standard might be: only authenticated users can submit feedback.

Ideally, every data-processing application should include suitable control mechanisms to prevent incorrect \cite{Martens:EMSE:2019} or missing data entry or processing~\cite{Guerra-GarciaPR20}, 
e.g., a predefined drop-down list of certain app versions
or a form template with a fixed structure. 
However, such forms might prevent users from sharing their feedback as they likely require more effort.   
Methods for data validation~\cite{SchelterLSCBG18,RedyukKMS21} are used to identify quality issues as early as possible and either fix them instantly or report them to the responsible data steward (e.g.~a product forum moderator or a support team). 
Especially when several applications use the same data,
installing a central data quality control is desirable.
For this reason, traditional relational database management systems offer data stewards the option of defining application-specific integrity constraints, which get monitored by the system~\cite{GarciaMolina2009}.  
Constraints could, e.g., check whether the information provided in the feedback matches that of the implicit feedback collected by a logger.

\subsection{Data Quality Improvement}
\label{sec:DQ_Improvement}
The main reasons for reduced data quality include 
missing, erroneous, inconsistent, ambiguous, incorrectly formatted, and outdated values~\cite{RahmD00,KimCHKL03,VisengeriyevaA20}.
Accordingly, data cleaning algorithms are developed to
impute missing values, find and fix data errors, 
detect and remove inconsistencies, and format data values consistently~\cite{IlyasC19,MahdaviAFMOS019}. Similar efforts have been made in the area of requirements engineering as well \cite{Montgomery:REJ:22}.
Another serious data quality problem is duplicates, 
i.e., different data records that refer to the same feedback entry 
(e.g.~a user reports the same bug in several channels)
or different feedback entries that refer to the same software issue
(e.g.~several users report the same bug which  affected them).
Detecting and merging duplicates is a non-trivial problem that has been highly discussed in the field of data management for several decades~\cite{Nau2010,Christen2012Book,2021Papadakis,LiLSWHT21}, as well as in the field of bug reporting for open source software \cite{Lueders:MSR:2022}.


The first step to improving the quality of a dataset is to learn more about its semantics.
For relational data, this includes semantic column and table types, such as \emph{firstname} for columns and \emph{product} for tables,
but also integrity constraints, such as functional dependencies, unique constraints, inclusion dependencies, or denial constraints.
Therefore, data profiling and schema extraction~\cite{2018Abedjan,SuharaL0ZDCT22,VisengeriyevaA20,KlettkeSS15,LbathBH21} are hot topics in current data management research.
In the case of user feedback, these methods can be applied to the metadata of the feedback, which were either additionally collected (e.g.~implicit feedback, demographics, channel dynamics, etc.) or previously extracted from the feedback itself. 

\section{Summary and Conclusion}
\label{sec:discussion}
User feedback is becoming more and more crucial for understanding user needs, monitoring software usage and its quality (particularly defects encountered in various contexts), as well as for planning and prioritising requirements. 
This chapter discussed how NLP and ML technology help deal with two major challenges for using feedback: the large quantity and the varying quality of feedback. 

Text classification, clustering, and summarisation techniques are the main steps to process feedback and cope with the quantity challenge. 
Classification approaches are nowadays fairly accurate due to the availability of labelled dataset and powerful transformer models, that can learn word context, deal with informal language, and heterogeneous vocabulary.
Recent LLMs also enables zero-shot classification with high accuracy.

There have also been advances in clustering and summarisation, however with limited success, as these are usually much harder tasks. 
Getting coherent clusters and groups of feedback with a short informative summary for developers can thus be challenging and still require several iteration with analyst-in-the-loop or redefining  the task as a classification task for popular  aspects discussed by the users. 
Therefore, one future area of research is to use visualisation and recommendation techniques to enable analysts to interact with the analysed feedback for a more effective retrieval of the desired information (e.g.~what are the most controversial feedback entries from disabled users to a certain app screen). 
Research on collecting and understanding implicit feedback \cite{Stanik2021} as well as user-developer discussion bots \cite{Martens:RE:19} may lead to promising results in this direction.

We think that coping with the feedback quality challenge is as important as coping with its quantity. 
Here, research is rather at its infancy phase. 
It is rather unclear what exactly good and bad feedback is and in which context. 
This should be clarified with empirical studies and possibly best practices and standards. 
Also tools to analyse and automatically rate the quality, as well as approaches to support users submit better feedback are limited and constitute a major future research direction. 
In this line, we think that researchers and practitioners can benefit from the area of data quality management.

%
%
%
\bibliographystyle{splncs04}
\bibliography{references}
\end{document}